\documentclass[AMA,STIX1COL]{WileyNJD-v2}

\usepackage{aastex_hack}

\articletype{Article Type}%

\received{26 April 2016}
\revised{6 June 2016}
\accepted{6 June 2016}

\begin{document}

\title{Evaluating a strategy for measuring deformations of the primary reflector of the Green Bank telescope using a terrestrial laser scanner}

\author[1]{Pedro Salas*}

\author[1]{Paul Marganian}

\author[1]{Joe Brandt}

\author[1]{Andrew Seymour}

\author[1]{John Shelton}

\author[1]{Nathan Sharp}

\author[1]{Laura Jensen}

\author[1]{Marty Bloss}

\author[1]{Carla Beaudet}

\author[1]{Dennis Egan}

\author[1]{Nathaniel Sizemore}

\author[1]{David T. Frayer}

\author[2]{Frederic R. Schwab}

\author[1]{Felix J. Lockman}

\authormark{Salas \textsc{et al}}

\address[1]{\orgname{Green Bank Observatory}, \orgaddress{Green Bank, \state{WV} 24944, \country{USA}}}

\address[2]{\orgname{National Radio Astronomy Observatory}, \orgaddress{Charlottesville, \state{VA} 22093, \country{USA}}}

\corres{*\email{psalas@nrao.edu}}

\abstract[Summary]{Astronomical observations in the molecule rich $3$~mm window using large reflector antennas provide a unique view of the Universe. To efficiently carry out these observations gravitational and thermal deformations have to be corrected. Terrestrial laser scanners have been used to measure the deformations in large reflector antennas due to gravity, but have not yet been used for measuring thermal deformations. In this work we investigate the use of a terrestrial laser scanner to measure thermal deformations on the primary reflector of the Green Bank Telescope (GBT). Our method involves the use of differential measurements to reduce the systematic effects of the terrestrial laser scanner. We use the active surface of the primary reflector of the GBT to validate our method and explore its limitations. We find that when using differential measurements it is possible to accurately measure deformations corresponding to different Zernike polynomials down to an amplitude of $60$~$\mu$m. The difference between the amplitudes of known deformations and those measured are $<140~\mu$m when the wind speed is $\lesssim2$~m~s$^{-1}$. From these differences we estimate that it should be possible to bring the surface error of the GBT down to $240\pm6~\mu$m. This suggests that using a commercial off-the-shelf terrestrial laser scanner it is possible to measure deformations induced by thermal gradients on a large parabolic reflector.
}

\maketitle

\footnotetext{\textbf{Abbreviations:} AS, active surface; FoV, field of view; GBT, Green Bank telescope; PTCS, precision telescope control system; TLS, terrestrial laser scanner}

\section{Introduction}\label{intro}

Astronomical observations at wavelengths $\lambda\leq1.5$~cm contain key information about planet and star formation, astrochemistry, galaxy evolution and cosmology. Large (diameter $\geq30$~m) single-dish radio telescopes operating at these wavelengths are unique in their ability to map large areas of the sky at good angular resolution in reasonable amounts of time. These large area maps are fundamental in linking small angular scales observed through interferometers, such as the Atacama Large Millimeter/submillimeter Array (ALMA), the NOrthern Extended Millimeter Array (NOEMA) and the Karl G. Jansky Very Large Array (VLA)\citep{Perley2011}, with their large scale astronomical environments.

For efficient observations, the root mean squared (rms) surface error of a reflector antenna should be $\lesssim\lambda/10$. To maintain the surface of a large reflector at this level, the effects of gravitational and thermal deformations must be mitigated and/or corrected. To account for gravitational deformations reflector antennas are built using homologous designs \citep{vonHoerner1967}.
With a homologous design, the effects of gravity are compensated by allowing the primary reflector to deform into a paraboloid of a different focal length that is a function only of elevation. Then, observations can be carried out by adjusting the position of the secondary reflector. Thermal deformations, those induced by gradients in the temperature of the telescope, can be mitigated by the use of thermal shielding or active temperature control. These involve maintaining and/or actively adjusting the temperature of the back up structure that supports the primary reflector to smooth out thermal gradients. Examples of radio telescopes using homologous designs and temperature control include the Institut de Radio Astronomie Millim\'etrique (IRAM) 30m telescope on Pico Veleta \citep{Baars1987} and the Nobeyama Radio Observatory (NRO) 45m radio telescope \citep{Ukita1994}. Another alternative is to correct gravitational and thermal deformations using a surface that can be adjusted in real time: an active surface (AS). Examples of telescopes using an AS are the Green Bank Telescope (GBT) \citep{Prestage2009}, whose individual surface panels are mounted at their corners on precision actuators whose length can be changed remotely in just a few seconds, the Large Millimeter Telescope (LMT) \citep{Hughes2020}, which will also incorporate a ventilation system for active temperature control of its backup structure \citep{Schloerb2020}, the Sardinia Radio Telescope (SRT) \citep{Bolli2015,Prandoni2017}, and the TianMa Radio Telescope, which only observes at wavelengths $\geq7$~mm. In this work we focus on the measurement of deformations of the primary reflector of the GBT.

The GBT has a $100$~m diameter primary reflector and operates at wavelengths between $88$~cm and $2.6$~mm with almost continuous wavelength coverage. This makes it the largest single-dish telescope operating in the molecule-rich atmospheric window that defines the $3$~mm spectral band. Observations at the shortest wavelengths ($\lambda\leq1.5$~cm) are made possible only through the AS on its primary reflector. The AS consists of $2004$ panels mounted on $2209$ actuators that are able to position the corners of the panels with a precision of $\pm25~\mu$m and a repeatability of $\pm50~\mu$m \citep{Prestage2009}.

Several corrections are used during GBT operations: (i) an elevation-dependent finite element model (FEM) to describe the effects of gravity, (ii) a focus tracking model that compensates for the change in the focal length of the primary reflector as a function of elevation, (iii) a Gravity-Zernike model that adds empirically determined corrections to the FEM, (iv) dynamic corrections to the position of the secondary reflector to compensate for the thermal flexing of the telescope, (v) a pointing model, (vi) holography to determine the actuator zero-points \citep{Hunter2011} and (vii) out-of-focus (OOF) holography \citep{Nikolic2007b} to correct for thermal deformations and any residual errors of the FEM and Gravity-Zernike model. Corrections (ii), (iv), (v) are collimation corrections, and (i), (iii), (vi), and (vii) are corrections applied to the primary reflector. The latest Gravity-Zernike model was derived from observations taken during the winter of $2013$ \citep{Maddalena2014}, and it makes the gain curve of the telescope almost independent of its elevation \citep{Frayer2018}. During the night, and under good weather conditions, it is possible to bring the surface error of the telescope to $\leq240~\mu$m rms using OOF holography. The continuous improvement of these models and the development of new measurement techniques is part of the precision telescope control system (PTCS) project.

For the GBT, the major unresolved source of surface error arises from deformations of its primary reflector caused by differential heating of the backup structure. These deformations can change on time scales of less than an hour, and, as they depend in a complex way upon the orientation of the telescope with respect to the sky, ground and the Sun, and to local weather conditions (e.g., wind and cloud cover), they cannot be easily modeled. Hence, their effects must be measured directly. The method currently used by the GBT to measure these deformations is OOF holography \citep{Nikolic2007a,Nikolic2007b}. OOF holography requires observations of a bright point-like source to map the telescope's beam with the receivers in-and-out of focus, a procedure that takes more than $10$~minutes in addition to the time it takes to slew to the bright point-like source. In practice this means that during the day weather induced deformations change on a timescale comparable to the time required to obtain the beam maps. This breaks the assumption of OOF holography that the surface remains stable during the observations, resulting in a degraded telescope gain for $3$~mm day time observations. Thus, a faster measurement method is required to enable efficient day time $3$~mm observations with the GBT. In this work we investigate the use of a terrestrial laser scanner (TLS) to measure deformations of the GBT's primary reflector caused by thermal gradients, part of the Laser Antenna Surface Scanning Instrument (LASSI) project.

A TLS is an active sensor that emits laser signals to calculate distances based on the time-of-flight of the returned laser pulses \citep{Vosselman2010}. In recent years, commercial TLS have been used to characterize the optics of radio telescopes \citep{Holst2015,Holst2019}. For example, Holst et al. have measured the effects of gravity on the focus of the primary reflector of the Onsala space observatory 20m radio telescope \citep{Holst2019} and of the 100m Effelsberg radio telescope \citep{Holst2015}. They were also able to identify large (${\approx}1$~mm) displacements of some of the panels on the primary reflector of the 100-m telescope \citep{Holst2015}. Motivated by this work and the potential of using a TLS for real-time telescope metrology, we have been investigating the use of a TLS to measure deformations an order of magnitude smaller (${\approx}100~\mu$m) than those previously identified.

The use of a TLS to measure thermal deformations has advantages and disadvantages. The potential advantages of using a TLS, over OOF holography, are that, (i) the TLS can scan the primary reflector in ${\approx}2$~minutes, while OOF holography takes more than $10$~minutes, and (ii) the TLS can scan the telescope at any orientation, while OOF holography requires observations of a point-like source, which requires slewing the telescope. On the side of the disadvantages, the distance measurements made with a TLS have relatively large systematic errors when compared to the accuracy required for $3$~mm observations. These systematic errors of a TLS are set by its design, which involves the use of a laser source and the two faces of a rotating mirror to scan large volumes efficiently (see e.g., \citep{Wujanz2017}). As the axes of rotation of the mirror are not perfectly aligned with the reference system of the scanner, the measured angles and distances are offset from their assumed values resulting in large systematic offsets (${\sim}1$~mm\citep{Holst2017}) over the scanned volumes. Here we mitigate these systematic effects, introduced by the mirror misalignment, by using differential measurements. In addition, individual distance measurements have an uncertainty (range uncertainty) of ${\approx}0.5$~mm rms at a distance of $50$~m. However, since the deformations of the primary reflector of the GBT are typically smooth and have scales of a few meters, they are sampled by thousands of individual distance measurements, so the range uncertainty can be averaged down.

\section{Methods}

Here we describe the theory behind the distance measurements made by the TLS and the use of differences to suppress their systematics when we measure deformations. Then we describe the TLS used by the LASSI, how we process the measurements made by this TLS for deformation analysis and how these processed distance measurements are combined to measure deformations of the primary reflector of the GBT.

\subsection{LASSI}
\label{ssec:lassi}

Each distance measurement by the TLS towards the telescope's primary reflector is a combination of its ideal parabolic shape, $S_{\mathrm{dish}}$, deformations from the ideal shape, $S_{\mathrm{d}}$, and the systematics introduced by the scanner itself, $S_{\mathrm{TLS}}$,
\begin{equation}
 S_{\mathrm{LASSI}}=S_{\mathrm{dish}}+S_{\mathrm{d}}+S_{\mathrm{TLS}}.
\end{equation}

To eliminate the systematics introduced by the TLS we use the difference between a signal and a reference measurement, $\Delta S_{\mathrm{LASSI}}$, taken at different times.
As the calibration parameters of the TLS appear to remain constant for days, if the measurements are made with the telescope at the same elevation (thus eliminating gravitational effects) this is
\begin{equation}
 \Delta S_{\mathrm{LASSI}}=S_{\mathrm{d}}^{\rm{sig}}-S_{\mathrm{d}}^{\rm{ref}}+\Delta S_{\mathrm{TLS}},
 \label{eq:defmap}
\end{equation}
where $S_{\mathrm{d}}^{\rm{sig}}$ and $S_{\mathrm{d}}^{\rm{ref}}$ are the deformations present during the signal and reference measurements, respectively, and $\Delta S_{\mathrm{TLS}}$ is the difference between the scanner systematics when the two scans were acquired. If the scanner has warmed up \citep{Janssen2021}, the ambient conditions (e.g., temperature, pressure, visibility) are the same when the two scans were acquired and the location of the TLS has not changed between the two measurements, then $\Delta S_{\mathrm{TLS}}\approx0$. In this case the uncertainty in $\Delta S_{\mathrm{LASSI}}$ is related only to the stochastic term of the scanner noise, $\sigma_{\mathrm{TLS}}^{\mathrm{sto}}$, its range uncertainty, as the systematic effects should cancel out.

Ideally $S_{\mathrm{d}}^{\rm{ref}}=0$, in practice, however, the primary reflector of the GBT is never a perfect paraboloid.
This means that when we correct for $\Delta S_{\mathrm{LASSI}}$ the surface error will be
\begin{equation}
 \varepsilon=\sqrt{\sigma_{\mbox{ref}}^{2}+\sigma_{\mbox{LASSI}}^{2}},
 \label{eq:srms}
\end{equation}
where $\sigma_{\mbox{ref}}$ is the surface error during the reference measurement (e.g. if we scan the surface after using OOF holography $\sigma_{\mbox{ref}}\approx230~\mu$m \citep{Frayer2018}) and $\sigma_{\mbox{LASSI}}$ the surface error introduced due to the imperfect measurement of deformations on the primary reflector.

The AS characterizes deformations from the ideal parabolic shape using Zernike polynomials.
A general deformation can be written as
\begin{equation}
 Z=\sum_{i}^{n_{Z}} C_{i}Z_{i},
 \label{eq:zernike}
\end{equation}
where $Z_{i}$ are the Zernike polynomials, $C_{i}$ are coefficients describing the magnitude of each Zernike polynomial, and $n_{Z}$ is the number of polynomials required to describe the deformation.
To determine the values of $C_{i}$ from $\Delta S_{\mathrm{LASSI}}$ we solve Eq.~\ref{eq:zernike} as a linear problem with $Z=\Delta S_{\mathrm{LASSI}}$.

The wavefront error of a Zernike polynomial is
\begin{equation}
 \varepsilon^{2}(C_{i})=\sum_{i}^{n_{Z}}\left(\frac{C_{i}}{N_{n}^{m}}\right)^{2},
 \label{eq:waveerror}
\end{equation}
were $n$ and $m$ are the radial and azimuthal degrees of the Zernike polynomial, respectively, and $N_{n}^{m}=\sqrt{(2n+1)/(\delta_{m,0}+1)}$, with $\delta_{m,0}$ the Kronecker delta.

The GBT's AS uses $n_{Z}=36$ and its own ordering for the polynomials. In the AS nomenclature $Z_{1}$ is the piston term ($n=0$ and $m=0$), $Z_{2}$ is tip (tilt in the cross-elevation direction, $n=1$ and $m=-1$), $Z_{3}$ is tilt (tilt in the elevation direction, $n=1$ and $m=1$), $Z_{4}$ is vertical astigmatism ($n=2$ and $m=2$), $Z_{5}$ is defocus ($n=2$ and $m=0$), $Z_{6}$ is oblique astigmatism ($n=2$ and $m=-2$) and $Z_{13}$ is the primary spherical term ($n=4$ and $m=0$). Then, Eq.~\ref{eq:waveerror} shows that the non-pointing related term with the highest contribution to the surface error is the defocus term, followed by vertical and oblique astigmatism and then higher order terms.

\subsection{TLS measurements}

\begin{figure}[h]
  \begin{minipage}[b]{0.4\textwidth}
    \vspace{0pt}
    \includegraphics[width=\textwidth]{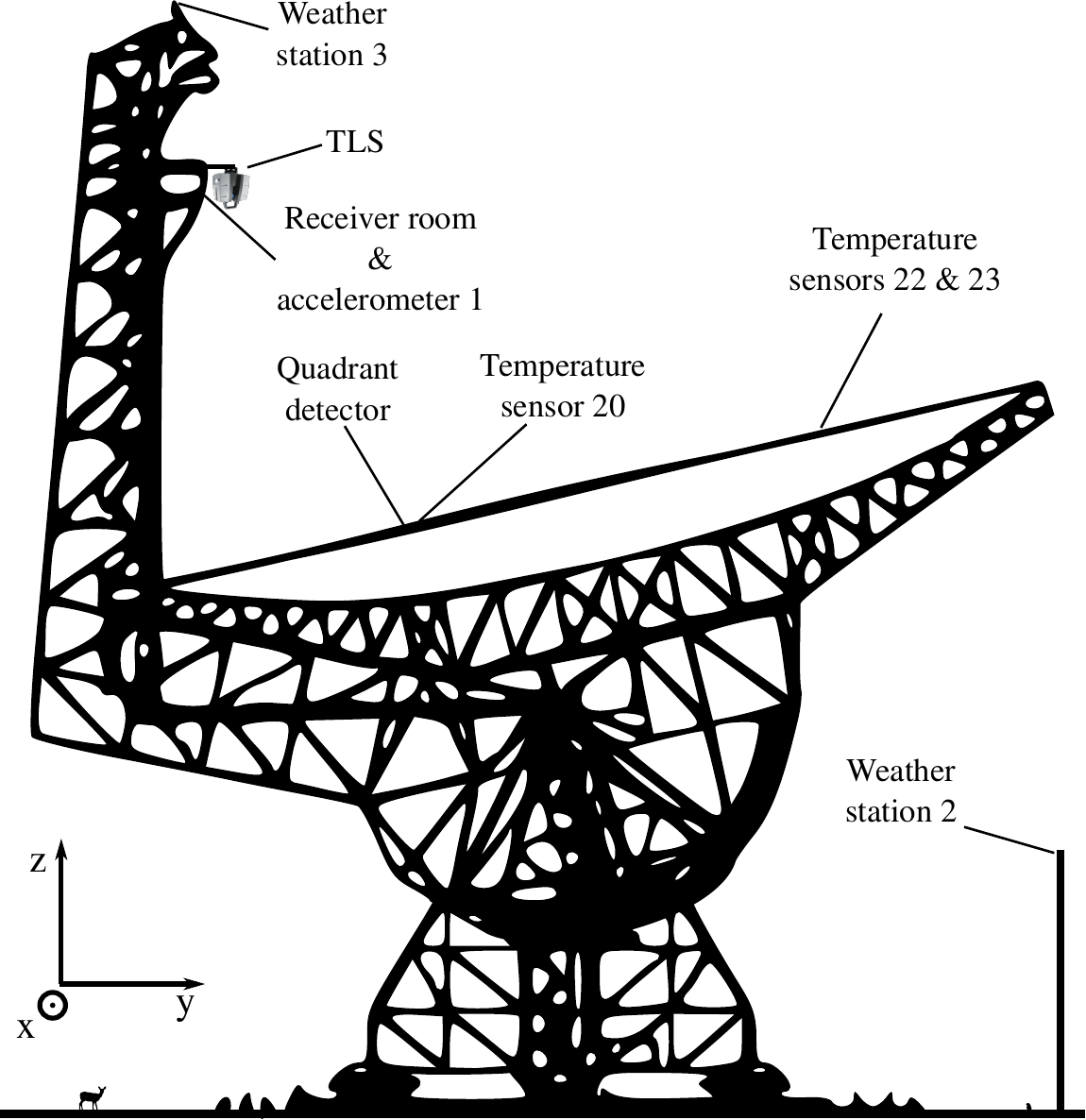}
  \end{minipage}\hfill
  \begin{minipage}[b]{0.55\textwidth}
    \vspace{0pt}
    \caption{Diagram showing the location of the TLS used by the LASSI, and some of the sensors installed on the GBT. The recordings from these sensors are used to compare against $\sigma_{\Delta S}$ on Figure~\ref{fig:external}. Temperature sensors $20$ and $22$ are located below the panels of the primary reflector, while $23$ is located on one of the beams of the back up structure. The coordinate system used by the LASSI is shown in the bottom left corner. Distances and sizes are not to scale.
         \label{fig:gbtdia}}
  \end{minipage}
\end{figure}

For the experiments presented in this work, we used a Leica ScanStation P40 TLS. The Leica ScanStation P40 has a field of view (FoV) of $360^{\circ}$ in azimuth and $290^{\circ}$ in elevation (to avoid measuring its own base), a maximum range of $200$~m, and, for mechanical reasons, it should be operated at an angle of $\leq10^{\circ}$ with respect to its gravity vector. For these reasons, to completely scan the GBT's primary reflector the TLS is mounted upside down on a beam extending out from the top of the receiver cabin, as shown in Figure~\ref{fig:gbtdia}. From this position the primary reflector is completely visible to the TLS when scanning with a $360^{\circ}$ FoV in azimuth. For LASSI we operate the TLS in speed mode with a resolution of $63$~mm at a distance of $100$~m. A scan of the full FoV with these settings takes $50$~s and produces $\approx20\times10^{6}$ distance measurements, which are collectively referred to as a \textquotedblleft point cloud".

\subsection{Point cloud processing}

\begin{figure*}[h]
 \includegraphics[width=\textwidth]{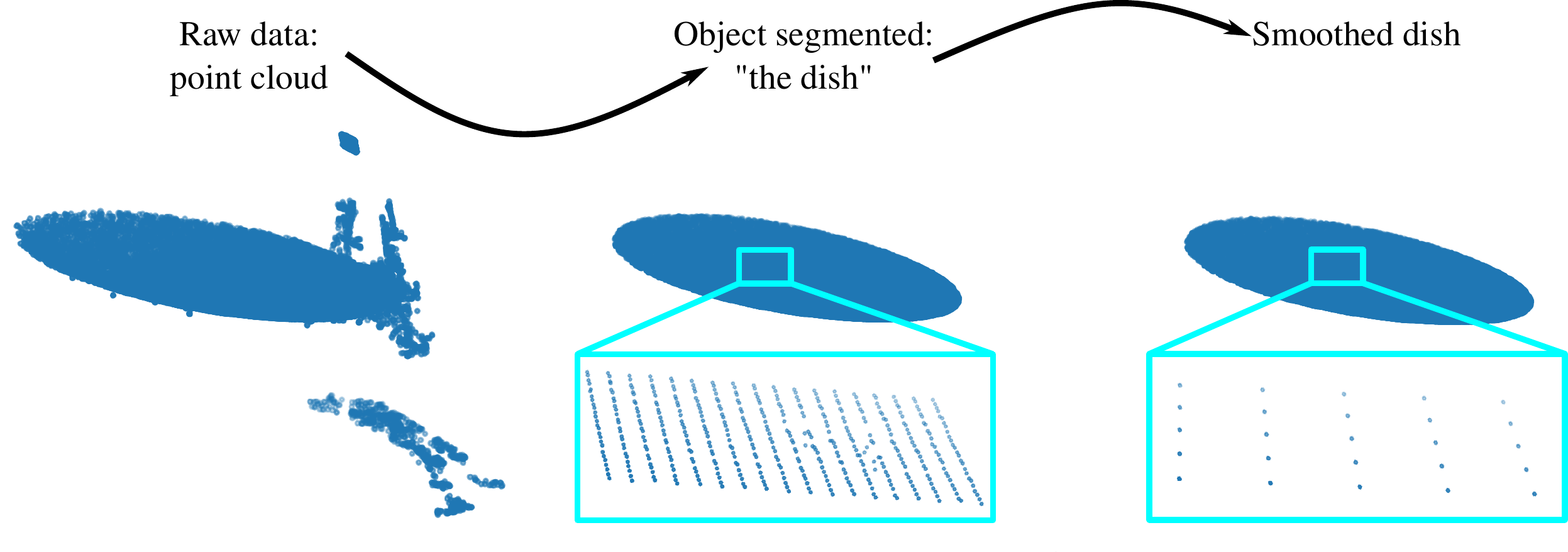}
 \caption{Visual representation of the object segmentation and smoothing steps during the point cloud processing.
          \label{fig:pipeline}}
\end{figure*}

The observed point cloud is processed to optimize it for deformation analysis. This processing includes the following steps; object segmentation, smoothing, gridding, masking and registration, which we describe below. A summary of the first two processing steps is presented in Figure~\ref{fig:pipeline}. Currently, the total time required to process a LASSI scan is $\approx6$~minutes, including the time required to acquire the data and transfer it to the data processing server.

\subsubsection{Object segmentation}
The purpose of object segmentation is to remove unwanted objects from the point cloud (e.g., feed arm, surrounding terrain, receiver cabin), isolating the primary reflector. Since the location of the TLS with respect to the primary reflector is known we use distances to remove the feed arm, receiver cabin and surrounding terrain. The feed arm and receiver cabin are within $49$~m from the TLS, while the ground is always farther than $110$~m. After removing the feed arm, receiver cabin and surrounding terrain we are left with approximately half of the original point cloud, and the majority of it samples the primary reflector. To remove any additional measurements which are not part of the surface of the primary reflector we fit a paraboloid to the remaining point cloud, and use the residuals between the best fit paraboloid and the point cloud to remove any outliers. For the fit, we use $1\%$ of the range measurements left after removing the feed arm, receiver cabin and ground to speed up the fitting process. This $1\%$ is randomly selected.

\subsubsection{Smoothing}
To reduce the number of range measurements in the point cloud we smooth the data in spherical coordinates using a Gaussian smoothing kernel with a FWHM of $0.001$~radians in the elevation direction and $0.001\cos(\mathrm{El})$~radians in the azimuth direction, so every smoothed distance measurement samples an equal solid angle. This produces a smoothed point cloud evenly sampled in spherical coordinates on a $512\times512$ grid, which samples the GBT surface every $\approx0.15$~m. Smoothing reduces the range uncertainty by a factor $\sqrt{n}$, with $n$ the number of measurements that are averaged together. This smoothing is computational taxing and is accelerated by using GPU processing. This step is performed to reduce the data volume while keeping the underlying information about the point cloud, and since we aim to process the point clouds as fast as possible to produce near real-time results.

\subsubsection{Gridding}
The smoothed point cloud is unevenly sampled in Cartesian coordinates. Since estimating the parameters of a surface using an unevenly sampled point cloud can lead to biases \cite{Holst2014,Holst2015}, we grid the data into an evenly sampled Cartesian grid using a linear interpolation to minimize these biases. The spacing between points in the Cartesian grid is $\approx0.2$~m, and its limits are determined from the extremes of the point cloud coordinates. This implies that the grid of each point cloud is independent.

\subsubsection{Masking}\label{sssec:masking}
Occasionally some of the 2209 actuators on the AS of the GBT are in need of repair which can result in some of the panels showing displacements $\sim1$~mm. Additionally, there are gaps and holes between the panels. The gaps avoid friction between panels and the holes give reference retroreflectors a clear view of the receiver cabin. The retroreflectors were installed as part of a photogrammetry system, used for the initial calibration of the panel and actuator positions, and are located $\approx5$~mm below the primary's surface. The highly reflective surface of the retroreflectors results in a large range uncertainty when their positions are measured using a TLS, hence they are not properly subtracted in the difference between scans and have to be masked.

To mitigate the effect of faulty actuators, gaps and holes during the measurement of deformations these are masked using distance and statistical thresholds. A first mask is derived by fitting a paraboloid to the smoothed gridded data. Residuals between the best fit paraboloid and the point cloud of the surface of the primary reflector which are more than $5\sigma$ away from the mean are masked. This masking usually removes range measurements at the edges of the primary reflector that were not appropriately captured during the object segmentation step, panels with large displacements, and the hole used by the quadrant detector. A second mask is derived for each row of the gridded data independently by fitting a third order polynomial to it using random sample consensus \citep[RANSAC, e.g,][]{Fischler1981}. RANSAC iterates over subsets of the data to find the best fit model, and in the process it also identifies outliers which are then masked.

\subsubsection{Registration}\label{sssec:align}
The smoothed, gridded and masked point clouds are registered to a common reference frame. The reference frame is set by a reference paraboloid
\begin{equation}
 z=\frac{x^{2}+y^{2}}{4f},
\end{equation}
where $f$ is the focal length of the paraboloid. To register the point clouds we include two rotations around the $x$ and $z$ axes, and three translations, so that the point cloud of the primary reflector is aligned with the reference paraboloid. The rotation along the $z$ axis (parallel to the gravity vector of the TLS) is pre-determined by the relative orientation of the coordinate system of the scanner and the coordinate system of the primary reflector. The rotation along the $x$ axis is determined from the elevation of the GBT as read from the encoders in its elevation drive. The translations are determined through a least squares fit of the smoothed and masked point cloud to the reference paraboloid. Additionally, we include $Z_{4}$, $Z_{5}$ and $Z_{6}$ during the least squares fit. Including these terms is important when we are registering scans where the primary reflector is deformed, as not including them results in additional shifts due to the degeneracy between the shape of the paraboloid and these deformations. The translations and rotation along the $x$ axis are derived for each scan independently, and are applied before performing the deformation analysis.

\subsection{Deformation analysis}

Once we have two smoothed, gridded, masked and registered point clouds, one for the reference scan and one for the signal scan, we proceed with the deformation analysis. The first step is to grid the point clouds to a common grid, so they sample the same points over the primary reflector. This is necessary since the initial gridding of the point clouds is done independently, so they do not necessarily sample the same points. During the gridding we use the masks derived for the individual scans when setting the grid limits. After gridding the individual masks are discarded. Then, we subtract the two gridded point clouds, Eq.~\ref{eq:defmap}, to produce a deformation map, and use Eq.~\ref{eq:zernike} to describe it in terms of Zernike polynomials. The coefficients describing the deformation map are saved to disk, along with the rest of the data for the observing session.
Before decomposing the deformation map into Zernike polynomials we derive a new mask based on the statistical properties of the deformation map. Any pixels that have values outside $\pm2\sigma$, with $\sigma$ the standard deviation of the deformation map, are not considered during the Zernike decomposition.

\section{Experiments with LASSI}\label{sec:obs}

In order to validate the measurement concept and to understand the different factors that influence the accuracy of LASSI we have taken scans under different conditions and times. A summary of the observations is presented in Table~\ref{tab:obs}. During all the observations the telescope was at an elevation of $77.8^{\circ}$, its access position. When we had the AS in its reference state, the actuators were set to their zero positions. When we had the AS in a signal state, we modified the actuator positions to introduce a deformation of the surface in the form of a Zernike polynomial. When doing mixed experiments we alternated between reference and signal states. For each state of the AS we performed a scan using the TLS.

\begin{center}
\begin{table*}[h]%
\caption{LASSI observations on the GBT.\label{tab:obs}}
\centering
\begin{tabular*}{500pt}{@{\extracolsep\fill}llccc@{\extracolsep\fill}}
\toprule
\textbf{Start time (UTC)} & \textbf{End time (UTC)}  & \textbf{AS} & \textbf{TLS mount}  \\
\midrule
June 11, 2019 15:00       & June 12, 2019 16:00      & Reference   & wedged \\
March 15, 2020 02:20      & March 16, 2020 10:00     & Mixed       & flat mount \\
\bottomrule                                             
\end{tabular*}
\end{table*}
\end{center}

For the scans taken during June, $2019$ the AS was kept on its reference state and the TLS was continuously scanning while mounted with a wedge under it, such that it is at angle of $10^{\circ}$ pointing away from the primary reflector. Using this wedge the TLS can fully map the primary reflector using an azimuthal FoV of $180^{\circ}$. During this experiments the point clouds were processed offline, so that the time between scans corresponds to the time it takes the TLS to acquire the point cloud, process it and transfer it to the observatory network. For the temporary setup used, this time is $\approx3$~minutes.

For the scans taken during March, $2020$ the TLS was on a flat and level mount, requiring a $360^{\circ}$ azimuthal FoV to fully capture the primary reflector. For the deformations we set all the $C_{i}$ to zero, except for one, and change the non-zero coefficient each time the AS is set to a signal state. We used $C_{4}=500,150,50~\mu$m, $C_{7}=500,150,50~\mu$m and $C_{13}=500,150,50~\mu$m. We looped over all the deformations for the duration of the observations. Durin this experiment the point clouds were processed in real time. This adds $\approx4.5$~minutes to the time between scans, for a total time between scans of $\approx7.5$~minutes.

For both experiments the telescope remains at a fixed elevation, so that the TLS is roughly stationary with respect to the surface of the primary relfector. However, since the telescope is not completely rigid, there will be small changes in their relative positions. Additionally, since the scans are acquired on short time scales (approximately $3$ or $7.5$~minutes), the systematics from the TLS remain roughly constant, so that $\Delta S_{\mathrm{TLS}}\approx0$ in Eq.~\ref{eq:defmap} after the initial warm up period.

\section{Results}

\subsection{Characterization of the TLS errors}

From the observations performed we characterize the performance of the TLS used by the LASSI, and how external factors, such as sunlight, affect it. First we analyze the stochastic errors and then the systematic errors from the TLS.

\begin{figure}[t]
 \begin{minipage}[t]{0.5\textwidth}
  \vspace{0pt}
  \includegraphics[width=\textwidth]{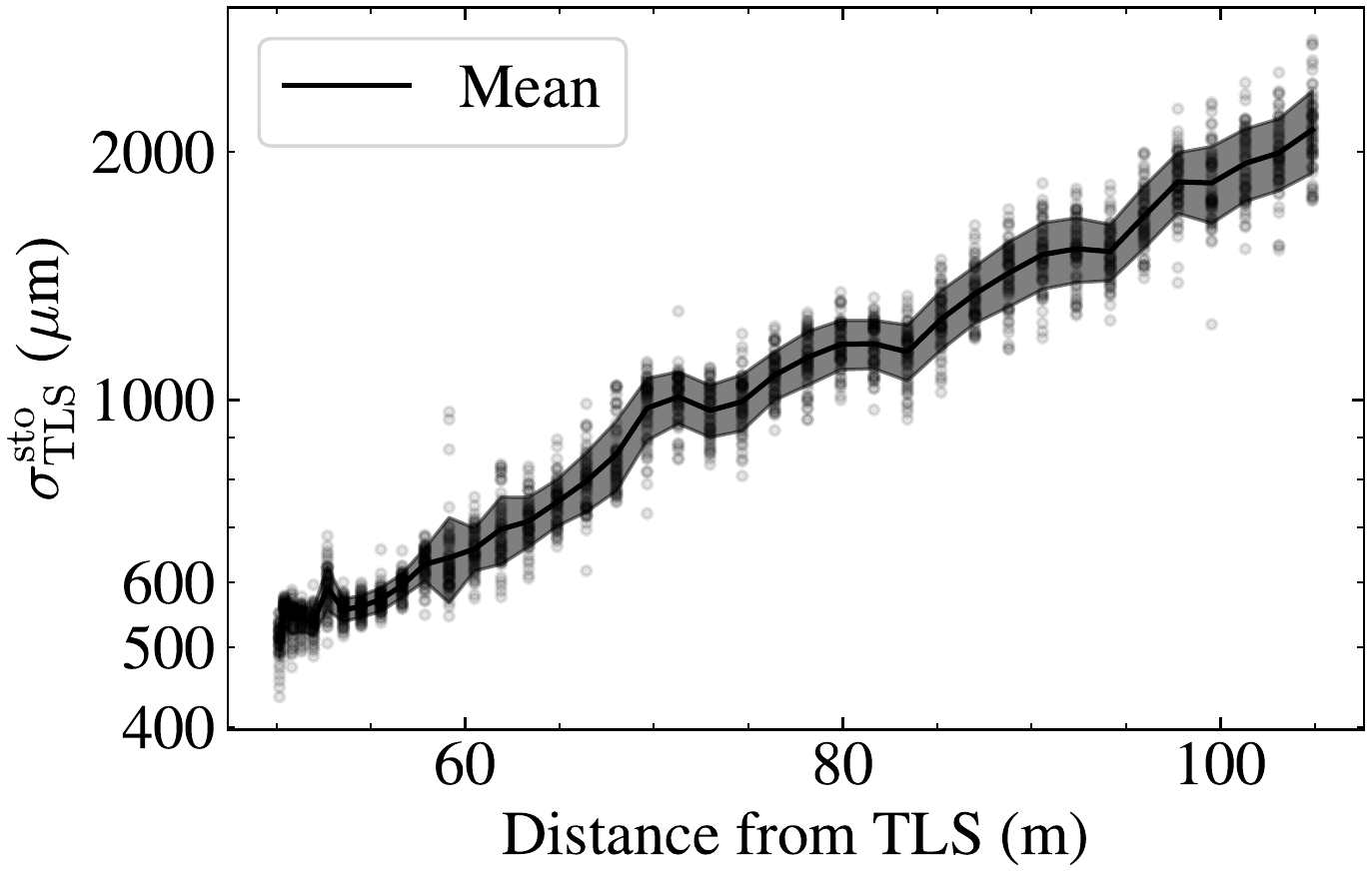}
 \end{minipage}\hfill
 \begin{minipage}[t]{0.4\textwidth}
  \vspace{0pt}
  \includegraphics[width=\textwidth]{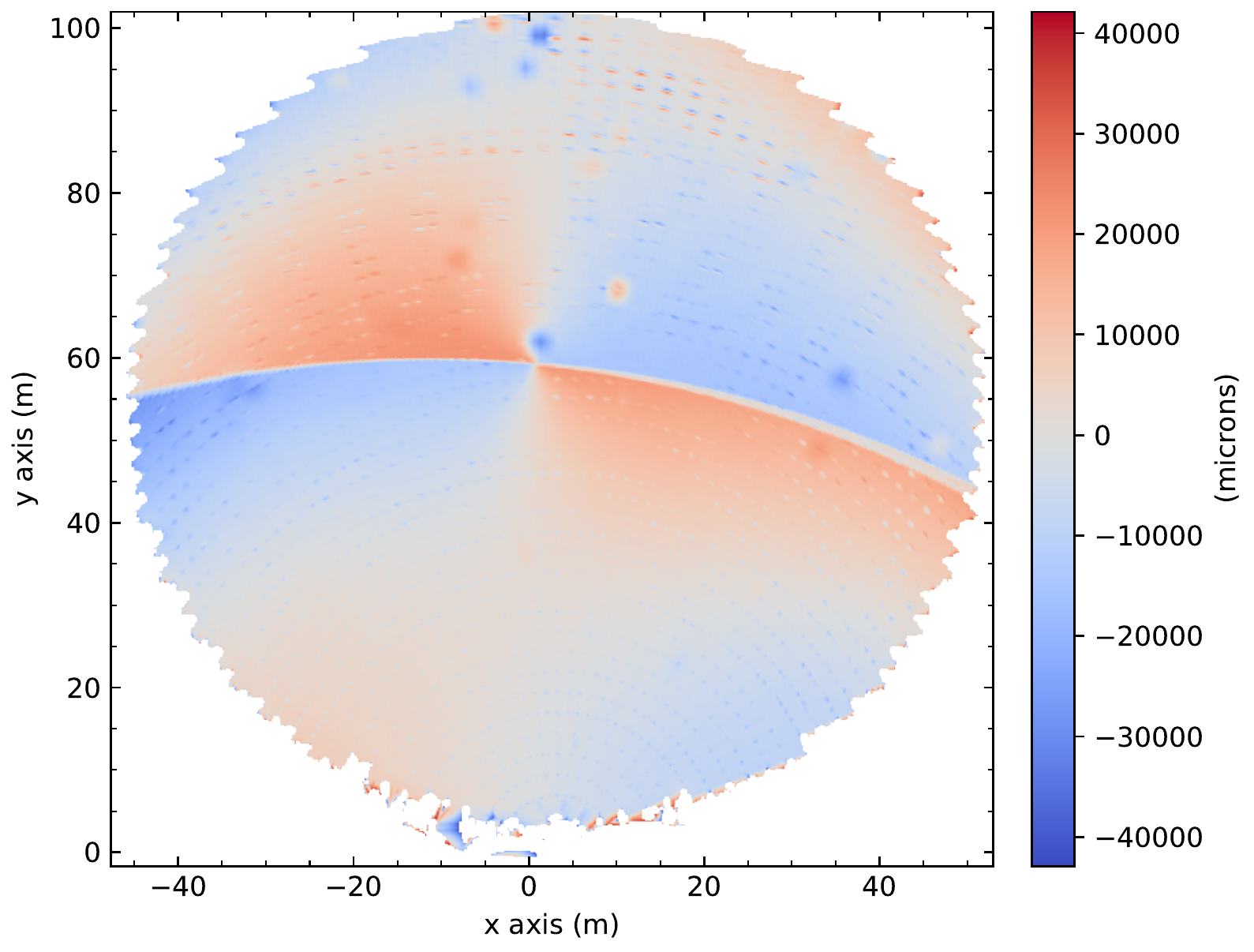}
 \end{minipage}
 \caption{\emph{Left}: Range uncertainty, $\sigma_{\mathrm{TLS}}^{\mathrm{sto}}$, for the TLS used by the LASSI as a function of distance. The gray dots represent individual measurements, the black line and the shaded region show the mean and the $1\sigma$ range around the mean, respectively. We use $61$ scans acquired in June 2019, to estimate the range uncertainty. To derive $\sigma_{\mathrm{TLS}}^{\mathrm{sto}}$ we measure the standard deviation of the residuals over individual panels. Since the panels have a surface rms of ${\approx70}~\mu$m they they do not contribute to the measured standard deviations which are dominated by the range uncertainty from the TLS.
         \emph{Right}: Residuals from the best fit paraboloid for a processed point cloud of the primary reflector of the GBT. The residuals show the systematics introduced by the misalignment of the optics of the TLS.
         \label{fig:tlsrms}}
\end{figure}

\subsubsection{TLS stochastic errors}

The stochastic errors on the distances measured by a TLS (range uncertainty) are a function of the intensity of the reflected laser beam \citep[e.g.,][]{Wujanz2017}, which in turn depends on other variables such as the distance to the target, the reflecting properties of the target, the angle of incidence of the beam to the target and the properties of the individual scanner. Here we use repeated measurements of the distance to the panels of the GBT's primary reflector to estimate the range uncertainty of the TLS used by the LASSI. In order to isolate the effect of the range uncertainty from the shape of the object being scanned we model each panel as a section of a paraboloid and measure the standard deviation of the distance measurements relative to this shape. Since the surface rms of individual panels is ${\approx}70~\mu$m they do not contribute significantly to the estimated range uncertainty. Additionally, since each panel covers a small solid angle (each panel has a surface of ${\approx}3.9$~m$^{2}$ \citep{Prestage2009}), the systematic effects introduced by the scanner are negligible in this analysis.

Figure~\ref{fig:tlsrms} shows the standard deviation of the distances measured by the TLS to the panels that make the surface of the primary reflector. The panels used to compute the range uncertainty lie on a straight line, from the vertex of the primary reflector to its far edge, which goes through the center of the primary reflector. We use $38$ out of the $45$ tiers of panels to avoid the edges of the primary reflector. Each data point in Figure~\ref{fig:tlsrms} corresponds to the standard deviation of the residuals between the best fit paraboloid that describes the shape of each panel, and the raw distance measurements from the scanner to the panel, for a single scan. We use $61$ scans from June $2019$, which sample the whole observation time. The dark line and the gray shaded areas in Figure~\ref{fig:tlsrms} show the mean of the range uncertainty and its $1\sigma$ ranges for each panel.

For a panel at a distance of $50$~m the range uncertainty is $517\pm28~\mu$m, and almost constant throughout the day. This is consistent with the range uncertainty reported by the manufacturer for a distance of $50$~m, $520~\mu$m, and indicates that the range uncertainty does not change significantly as the surface of the primary reflector is illuminated by the Sun. The range uncertainty increases with distance, $r$, as $\sigma^{\rm{sto}}_{\rm{TLS}}\propto r^{2}$, reaching a value of ${\approx}2.5$~mm at the far edge of the dish ($110$~m from the TLS). The scaled intensity of the reflected laser beam decreases with distance as $I\propto r^{-0.2}$. This is shallower than the relation expected from the radar range equation ($I\propto r^{-2}$) and is likely explained by distance-dependent amplification of the intensity by the TLS firmware.

When the raw data is smoothed approximately $100$ distance measurements are averaged together resulting on a range uncertainty per averaged distance measurement of ${\approx}62~\mu$m at $50$~m from the TLS, and larger towards the edges of the primary reflector where the intensity returned to the TLS is lower and the density of the point cloud is smaller.

\subsubsection{TLS systematic errors}

As discussed in the introduction, one of the limitations in the use of TLS arises from the misalignment of its spinning mirror relative to the axes of rotation of the mirror. This misalignment introduces large ($\sim1$~mm) systematic errors in the point cloud. Examples of these errors are presented in the right panel of Figure~\ref{fig:tlsrms}, where we show the residuals from the best fit paraboloid to a processed point cloud. The residuals reach values $\pm40$~mm, and are larger than the surface error of the primary reflector even if no corrections for thermal deformations are applied ($\varepsilon\approx500~\mu$m). Additionally, the portions of the dish scanned with the front and back faces of the TLS are also clearly distinguishable due to the discontinuity running almost horizontal at $y\approx55$~m. The prescence of these artifacts precludes the use of single scans with an uncalibrated TLS to measure deformations on the primary reflector of the GBT since they have large amplitudes and show up over large scales, hidding the lower level deformations the LASSI aims to measure.

\subsection{The influence of external effects on LASSI}

\begin{figure}[h]
\centerline{\includegraphics[width=\textwidth]{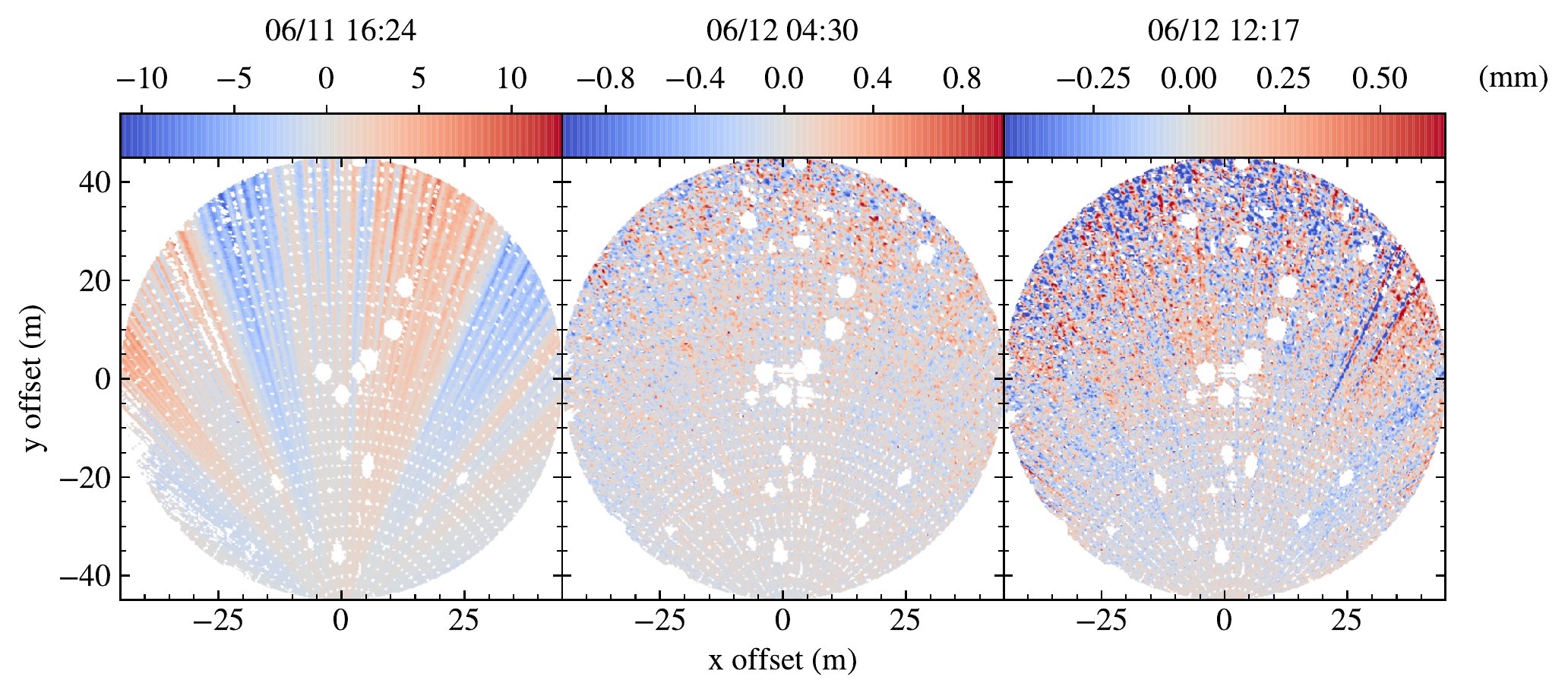}}
\caption{Exampled of deformation maps obtained during the scans taken in June 2019.
         Retroreflectors and panels with large residuals with respect to the paraboloid have been masked.
         These scans were taken prior to the summer maintenance period, when the AS' actuator zero points are calibrated using holography\citep{Hunter2011}, and faulty actuators replaced. 
         \label{fig:exdefmaps}}
\end{figure}

To understand how external factors (e.g., wind, temperature, sunlight) affect our ability to measure the surface using LASSI we compare the rms in the deformation maps, $\sigma_{\Delta S}=\sqrt{1/n_{x}n_{y}\sum\Delta S_{\mathrm{LASSI}}^{2}}$, to the recordings of the sensors installed on the GBT and around it. We use the June $2019$ scans for this comparison, during which time the AS remained fixed. The rms is measured on deformation maps computed from scans taken $3$~minutes apart, so that the effect of thermal distortions of the primary reflector is negligible. The analysis was also restricted to only the innermost $92\%$ of the reflector to minimize edge effects. Three examples of the deformation maps used to measure $\sigma_{\Delta S}$ are presented in Figure~\ref{fig:exdefmaps}. The location of some of the sensors used is shown in Figure~\ref{fig:gbtdia}. The sensors include $23$ thermistors installed across the telescope\footnote{The locations of the temperature sensors can be found in \href{https://safe.nrao.edu/wiki/pub/GB/PTCS/TemperatureSensorLocationDrawings/44002M002c.pdf}{https://safe.nrao.edu/wiki/pub/GB/PTCS/TemperatureSensorLocationDrawings/44002M002c.pdf}}, three weather stations with anemometers, four accelerometers (mounted on the receiver room, the left and right sides of the elevation bearing casting and close to the telescope's vertex), and a quadrant detector mounted below the surface of the primary reflector and illuminated from the feed arm so that it measures the relative displacement between the feed arm and the primary reflector. The quadrant detector and accelerometer $1$ have a sample rate of $10$~Hz, and the sensitivity of accelerometer $1$ is $\pm1.3\times10^{-4}$g.

\begin{figure}[!hb]
  \begin{minipage}[t]{0.47\textwidth}
  \vspace{0pt}
   \includegraphics[width=\textwidth]{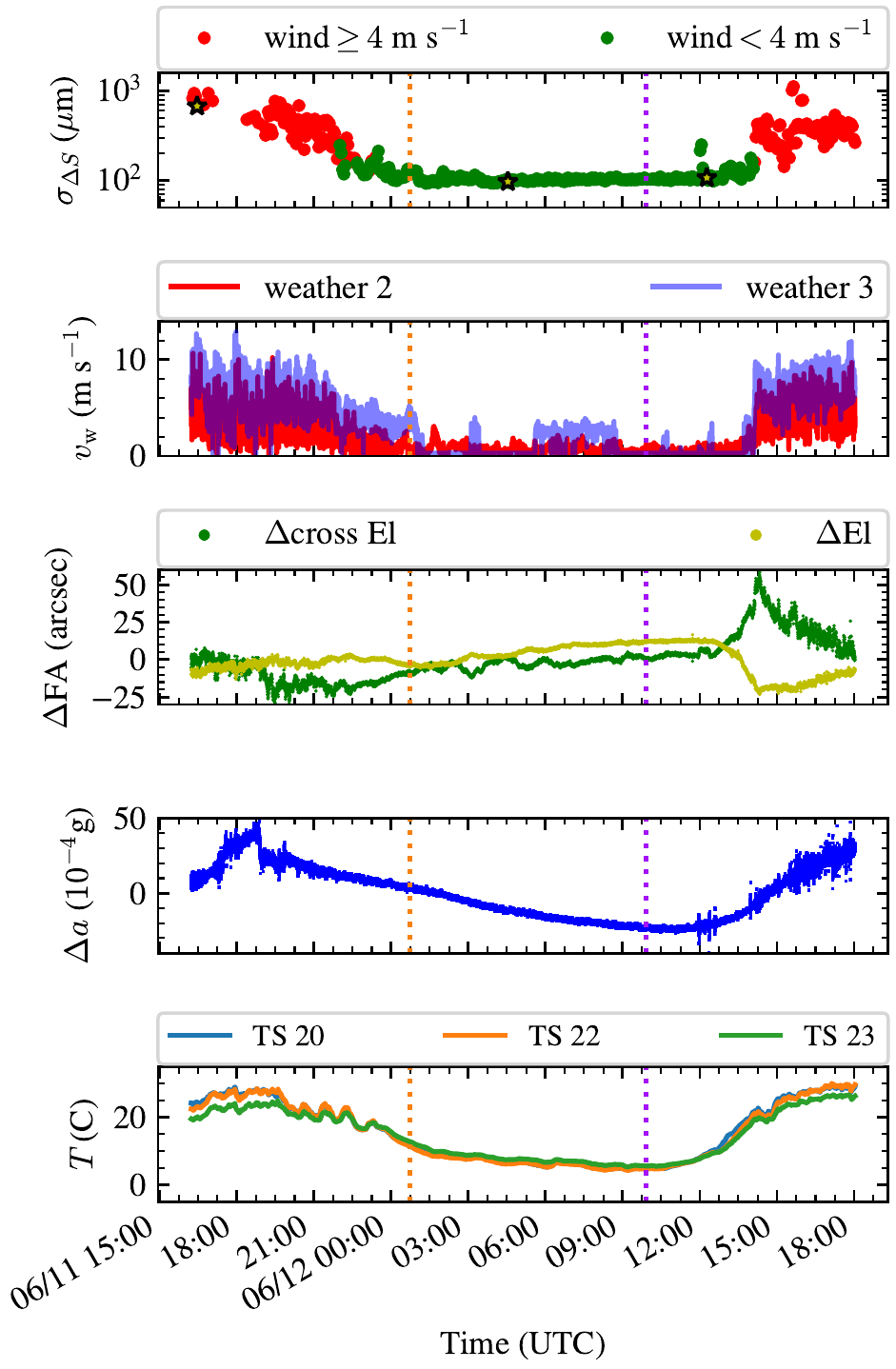}
  \end{minipage}\hfill
  \begin{minipage}[t]{0.47\textwidth}
  \vspace{0pt}
   \includegraphics[width=\textwidth]{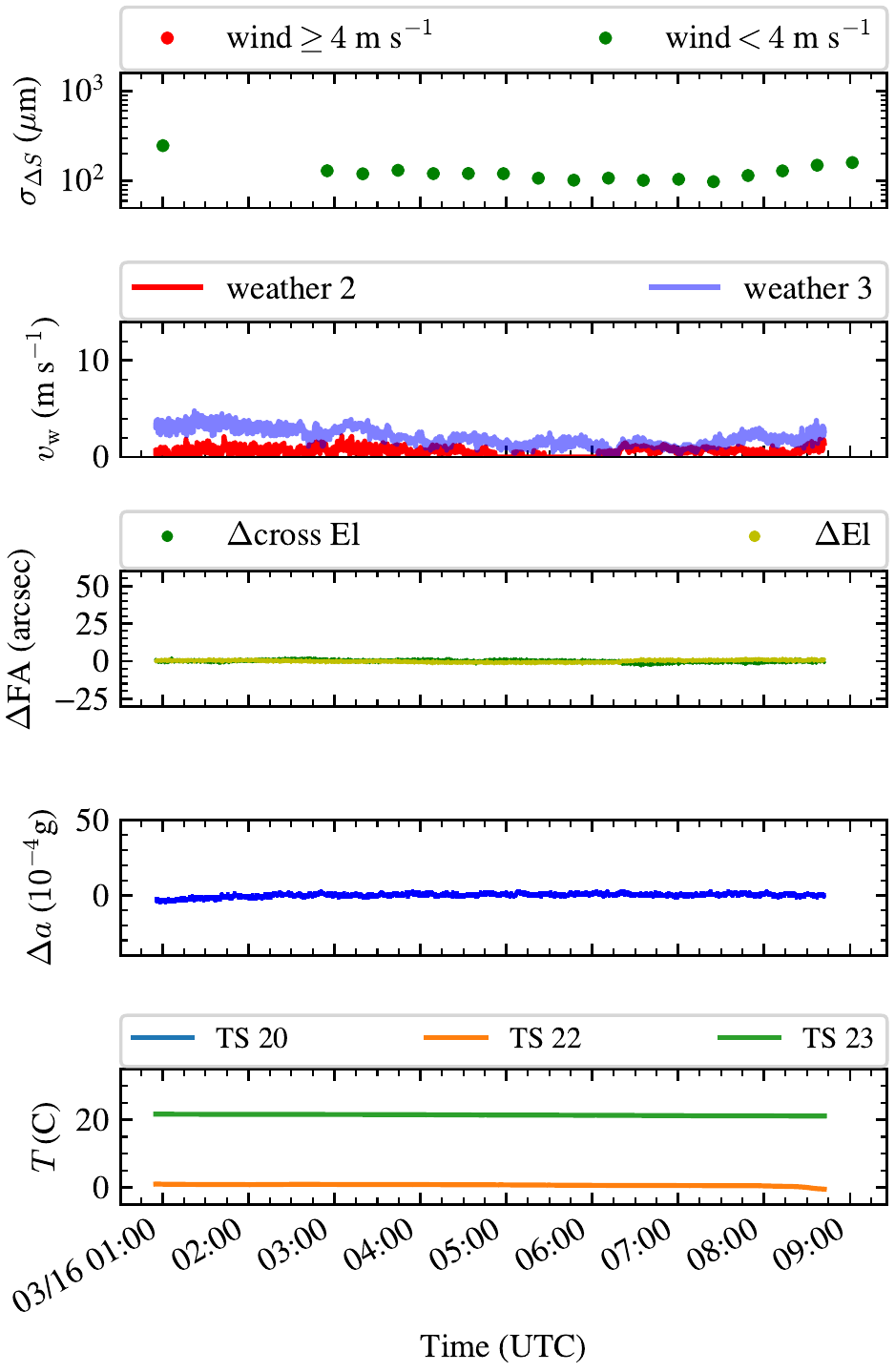}
  \end{minipage}
  \caption{\textit{Left}: Results from the $24$~hour scan performed during June 11 and 12, 2019. From top to bottom; $\sigma_{\Delta S}$, the rms of the surface deformation maps; wind speed recorded by two weather stations, weather 2 is located in the ground, weather 3 is located on the feed arm of the GBT; total displacements of the feed arm as recorded by the quadrant detector (QD); rms of the acceleration recorded by accelerometer 1 (located on top of the receiver cabin); temperature of the primary reflector at three different positions (BUS 36+000-PANEL, BUS 16+000 PANEL and BUS 36+000-BEAM, for more details see the \href{https://safe.nrao.edu/wiki/bin/view/GB/PTCS/TemperatureSensorDocumentation}{PTCS wiki}). The orange dotted line marks sunset (Sun's elevation $-50^{\prime}$), while the purple dotted line marks sunrise (Sun's elevation $-50^{\prime}$). The times at which the scans used to compute the deformation maps shown in Figure~\ref{fig:exdefmaps} where acquired are marked with yellow stars on the top panel.
           \textit{Right}: Same as left column but for the March 15, 2020 scans. The observations were carried out while the Sun was below the horizon. The readings of temperature sensor 23 are incorrect during this time period, and temperature sensor 20 was not recording data.
           The dates on the x axis are given as month/day hour:minutes.
           \label{fig:external}}
\end{figure}

We present the comparison between $\sigma_{\Delta S}$ and the readings of the different sensors in Figure~\ref{fig:external}. The left panel of Figure~\ref{fig:external} shows the results from the June $12$, 2019 scans. First we note that the wind speeds recorded by weather station $3$ are not reliable during this time period. In general we expect that the wind speed recorded by weather station $3$ will be larger than that recorded by weather station $2$ due to the altitude difference between the two stations, but we observe that the wind speeds recorded by weather station $2$ are larger than those from weather station $3$ for parts of the observation. For this reson we use the wind speeds from weather station 2 for the rest of the analysis of the June 15, 2019 scans.

The top panel of the left column shows that $\sigma_{\Delta S}$ is larger before sunset and after 10:30~am on June $12$, 2019. During these periods the wind speeds, the displacement of the feed arm relative to the primary reflector and the changes in the acceleration are the largest (second, middle and fourth rows on the left column of Figure~\ref{fig:external}, respectively). The correlation between the wind speed, the motion registered by the quadrant detector and the changes in acceleration can be understood in terms of the loading of the feed arm due to the wind \citep{Constantikes2003,Ries2009}. This leads us to hypothize that when the wind speed is above a certain threshold it moves the feed arm, where the scanner is located, relative to the primary reflector, the object being scanned, resulting in the larger rms in the deformations maps. Since the TLS requires approximately $50$~s to capture the point cloud, any relative movements between the feed arm and the primary reflector will result in artifacts in the deformation maps. These artifacts will not only increase the noise in the deformation maps, but also limit the ability of the LASSI to measure deformations of the primary reflector. Examples of the artifacts induced by the relative motion between the feed arm and primary reflector can be observed in the left panel of Figure~\ref{fig:exdefmaps} as rays originating from the location of the scanner (at $x\approx0$~m and $y\approx-45$~m). The scans used to produce this deformation map were obtained during a period of high wind speed as shown in Figure~\ref{fig:external} (first yellow star in the top left panel).

\begin{figure}[ht]
  \begin{minipage}[c]{0.75\textwidth}
  \vspace{0pt}
   \includegraphics[width=\textwidth]{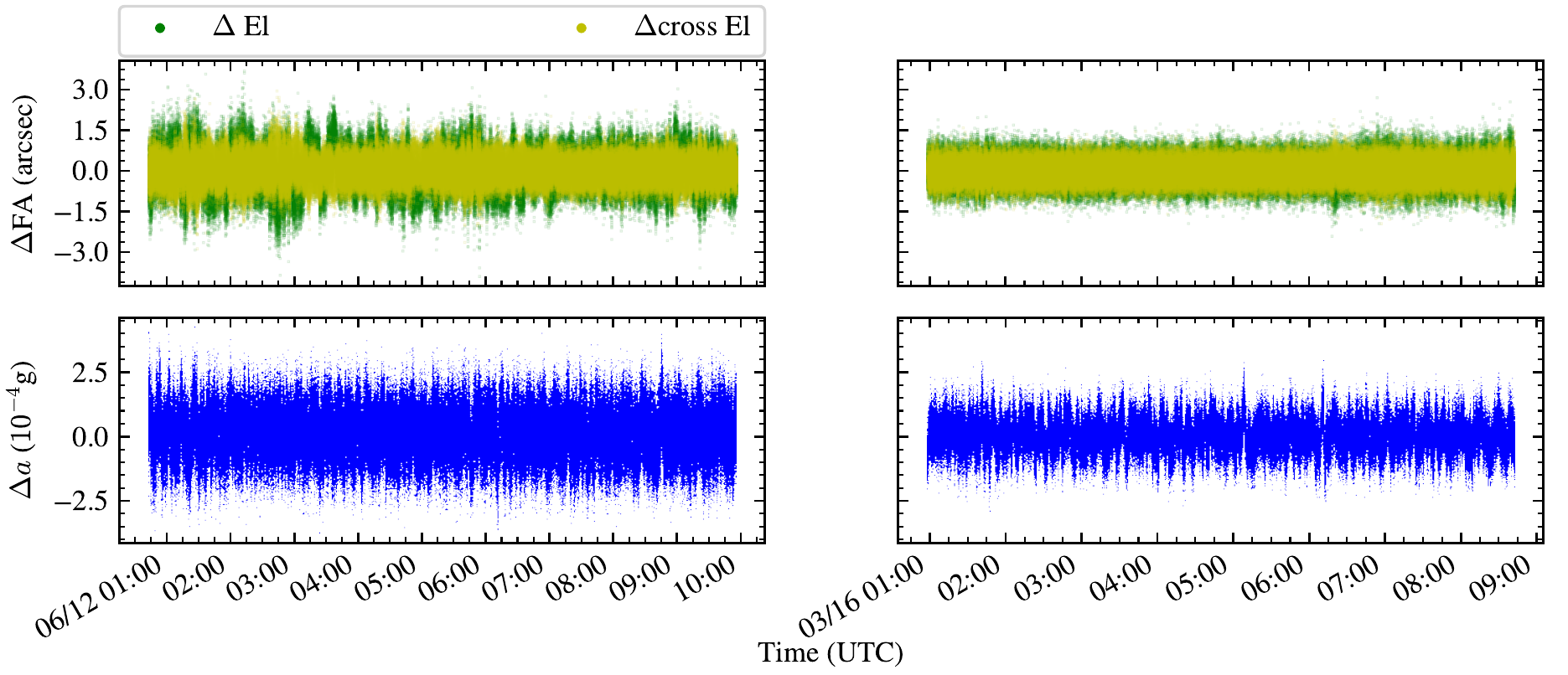}
  \end{minipage}\hfill
  \begin{minipage}[c]{0.25\textwidth}
  \vspace{0pt}
  \caption{Deflections of the feed arm relative to the primary reflector (top row) and acceleration registered at the roof of the receiver cabin (bottom row) for the June 2019 (left) and March 2020 scans (right). The readings have had a running average subtracted to remove changes on scales of $2.5$~minutes. The dates on the x axis are given as month/day hour:minutes.
           \label{fig:accqd}}
  \end{minipage}
\end{figure}

The right column of Figure~\ref{fig:external} shows $\sigma_{\Delta S}$ and the readings from the sensors being considered for the March 15, 2020 scans. To compute $\sigma_{\Delta S}$ we use deformation maps calculated from the difference of point clouds acquired while the AS was in its reference state. Thus, the time between the scans used to produce the deformation maps was $\approx14$~minutes. $\sigma_{\Delta S}$ has an average value of $127\pm34~\mu$m. In comparison, during June 2019, the night time average of $\sigma_{\Delta S}$ was $103\pm6~\mu$m. The values of $\sigma_{\Delta S}$ are consistent between the two nights. To compare the readings of the quadrant detector and accelerometer 1 during both nights we subtract the mean computed over a running box window from the time series. The box width is $2.5$~minutes. The mean subtracted readings for the two nights are shown in Figure~\ref{fig:accqd}. This comparison shows that the deflections of the feed arm relative to the primary reflector and the acceleration of the receiver cabin were larger during the night of June 12, 2019. This suggests that the movement of the feed arm relative to the primary reflector is not the only factor that determines $\sigma_{\Delta S}$. We will further investigate what other factors determine $\sigma_{\Delta S}$ in future work.

\subsection{Measuring deformations of the GBT's primary reflector}

To determine the accuracy with which we can measure deformations of the primary reflector we compare the measurements of $C_{i}$ from deformation maps to the deformations introduced through the AS.
For the March, $2020$ scans the AS was commanded to deform using Zernike polynomials with $C_{i}=50,\,150\mbox{ and }500~\mu$m for $i=4,\,7\mbox{ and }13$. Since the actuators of the AS have a precision of $\pm25~\mu$m and repeatability of $\pm50~\mu$m, we expect differences between the commanded deformations and the actual displacement of the actuators. To capture these differences, we used the relative displacements of the actuators as recorded by their encoders. From these relative displacements we reconstruct the deformations of the AS between the reference and signal scans. We find differences of $\pm50~\mu$m between the commands to the AS and the Zernike polynomials measured from the displacement of the actuators, $C_{i}^{\rm{AS}}$, consistent with the repeatability of the actuator positioning, however we note that $C_{i}^{\rm{AS}}$ is always larger than the commanded values. The deformation maps used to measure deformations are computed using the difference between scans when the AS was set to a signal state and a reference state.

\begin{figure}[h]
 \includegraphics[width=\textwidth]{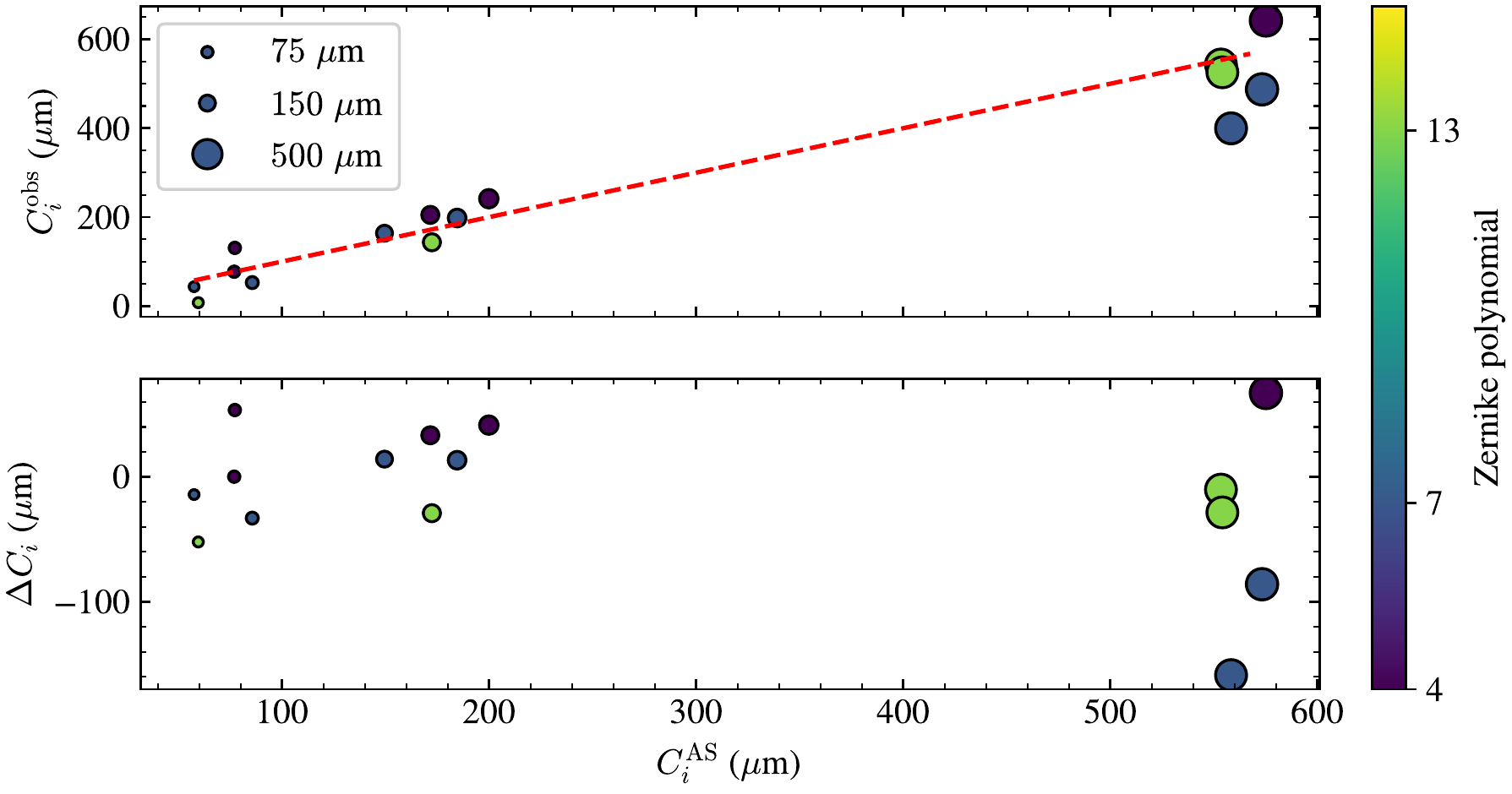}
 \caption{\textit{Top:} Zernike coefficients measured from surface deformation maps, $C_{i}^{\mathrm{obs}}$, as a function of the deformations input to the AS of the GBT, $C_{i}^{\mathrm{AS}}$.
         The red line shows a one-to-one relation, i.e., perfect measurement of the coefficients.
         \textit{Bottom:} Difference between the measured and input values. The color of each circle shows the Zernike polynomial and their sizes represent the coefficient value.
         The color of the circle shows the Zernike polynomial used to deform the primary reflector, and its size the amplitude of the deformation.
         \label{fig:zernikeinout}}
\end{figure}

In Figure~\ref{fig:zernikeinout} we show the difference between the Zernike polynomials measured from the deformation maps and those produced by the AS. There is a strong correlation between the two, showing that the LASSI is able to accurately measure the deformation of the AS. The difference between the measured values is $\leq140~\mu$m, with a mean and $1\sigma$ scatter of  $23\pm67~\mu$m. The scatter is comparable to the repeatability of the positioning of the actuators. We note that the largest differences between $C_{i}^{\mathrm{obs}}$ and $C_{i}^{\mathrm{AS}}$ are for $C_{7}=500~\mu$m, which is lower in $C_{i}^{\mathrm{obs}}$ than $C_{i}^{\mathrm{AS}}$.

In Figure~\ref{fig:zernikeinout} we compared the coefficients of Zernike polynomials which were explicitly set to non-zero values. However, even if we set $C_{i}=0$ the AS will deform as we are changing the positions of all the panels between scans and the encoders are not perfect. Additionally, the LASSI might measure additional deformations for other Zernike polynomials. To analyze these effects we use Eq.~\ref{eq:waveerror} for $4\leq i\leq36$, which captures any differences between the total deformation of the AS and the deformations measured by the LASSI. This is also an estimate of $\sigma_{\mathrm{LASSI}}$, as it shows how much the measurements made by the LASSI deviate from the deformations in the primary reflector. 

\begin{figure}[h]
  \begin{minipage}[t]{0.5\textwidth}
    \includegraphics[width=\textwidth]{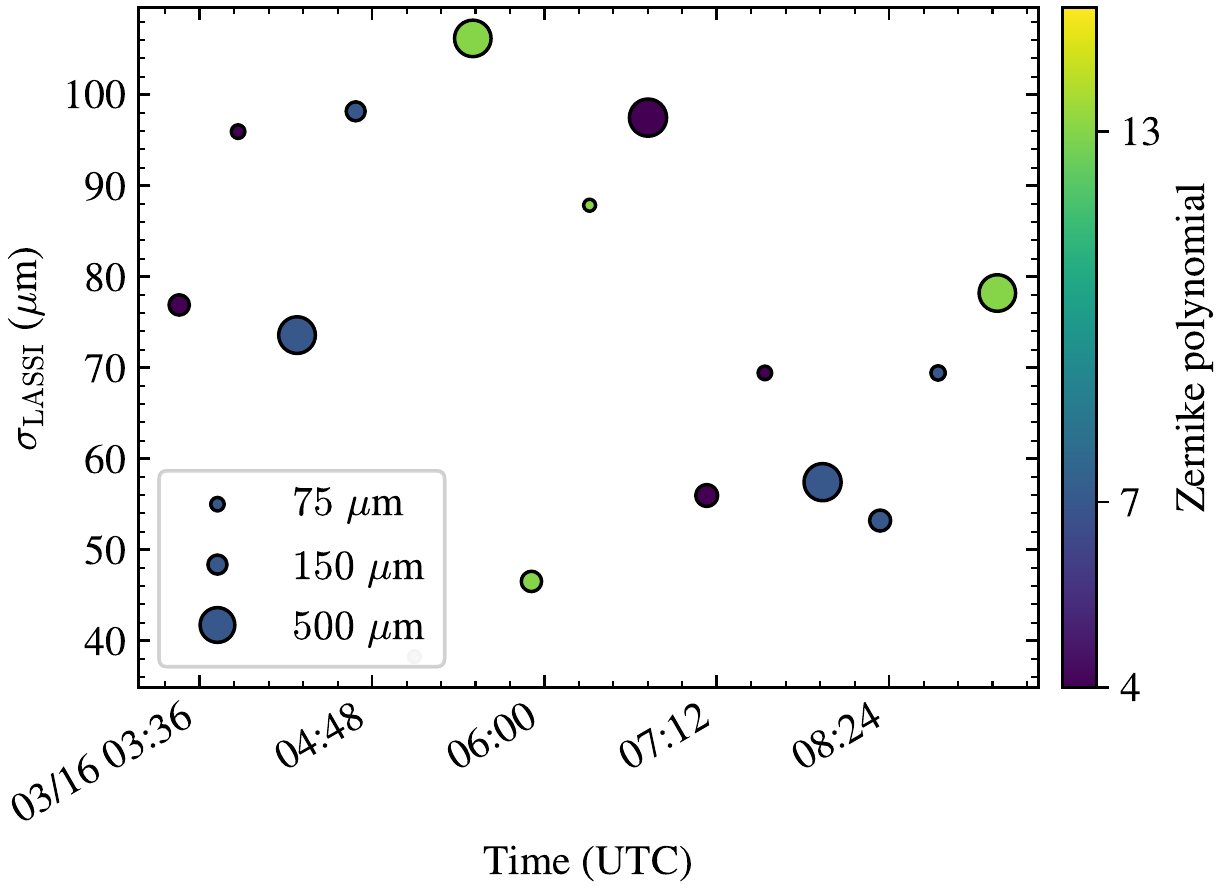}
  \end{minipage}\hfill
  \begin{minipage}[t]{0.5\textwidth}
   \includegraphics[width=\textwidth]{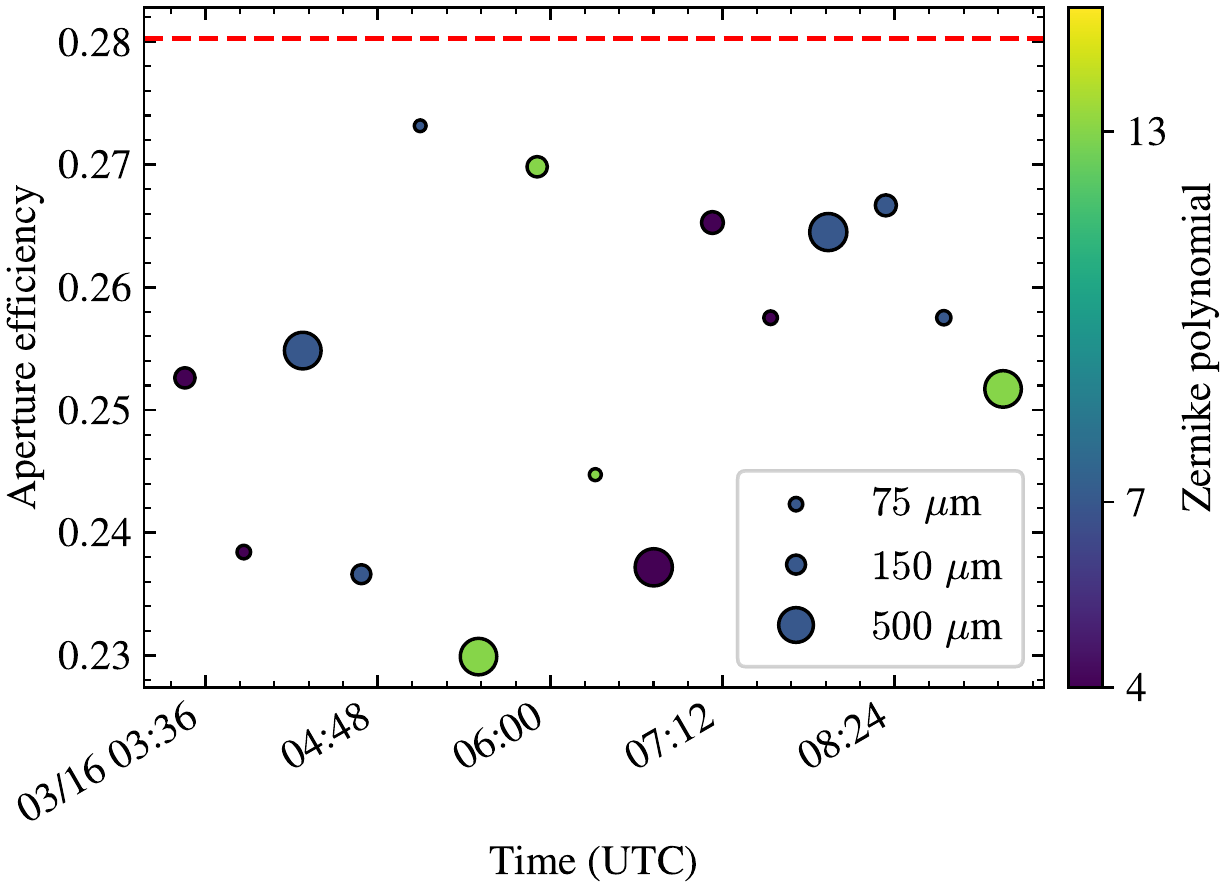}
  \end{minipage}
   \caption{\textit{Left}: $\sigma_{\mathrm{LASSI}}$ as a function of time for the March 14, 2020 experiments. $\sigma_{\mathrm{LASSI}}$ is estimated from the difference between $C_{i}^{\mathrm{obs}}$ and $C_{i}^{\mathrm{AS}}$, see text for details.
   \textit{Right}: Aperture efficiency at $3$~mm estimated from $\sigma_{\mathrm{LASSI}}$. To estimate the aperture efficiency we assumed an ideal surface with an rms error of $230~\mu$m and an aperture efficiency of $0.71$ at $21$~cm. The red dashed line shows the aperture efficiency for a surface rms of $230~\mu$m. The circles show the aperture efficiency when adding $\sigma_{\mathrm{LASSI}}$ to the surface error. The abscissa axis shows the time at which the point clouds used to measure $C_{i}^{\mathrm{obs}}$ were acquired.
   The color and size of the circles are the same as in Figure~\ref{fig:zernikeinout}.
         \label{fig:eta}}
\end{figure}

Our estimates of $\sigma_{\mathrm{LASSI}}$ are shown in the left panel of Figure~\ref{fig:eta}. The mean value is $\sigma_{\mathrm{LASSI}}=74\pm20~\mu$m. This is smaller than the difference between $C_{i}^{\mathrm{obs}}$ and $C_{i}^{\mathrm{AS}}$ because of how each Zernike polynomial contributes to the wavefront error (Eq.~\ref{eq:waveerror}). $\sigma_{\mathrm{LASSI}}$ shows no clear correlation with the polynomial used to deform the primary reflector or its amplitude, in contrast with $|\Delta C_{i}|$ which was larger for $C_{7}=500~\mu$m (Figure~\ref{fig:zernikeinout}). In general, $\sigma_{\mathrm{LASSI}}$ is dominated by the contribution from terms other than the deformation introduced, that is, if we deformed the surface using $C_{7}$, more than $63\%$ of the wavefront error comes from terms other than $C_{7}$. The major contribution to the wavefront error comes from $C_{4}$, $C_{5}$ and $C_{6}$, which account for $\approx65\%$ of the wavefront error. This highlights the importance of being able to accurately measure the low order Zernike polynomials. For $\sigma_{\mathrm{LASSI}}=74\pm20~\mu$m, the total surface error (Eq.~\ref{eq:srms}) is $\varepsilon=240\pm6~\mu$m. This is lower than the surface error in the absence of OOF holography corrections, $\varepsilon\approx300~\mu$m.

Given the differences between $C_{i}^{\mathrm{obs}}$ and $C_{i}^{\mathrm{AS}}$ of Figure~\ref{fig:zernikeinout} we can estimate the increase in the surface error and the decrease in aperture efficiency. The increase in the surface error is determined as $\varepsilon^{2}=\varepsilon_{0}^{2}+1/n_{x}n_{y}\sum Z^{2}(C_{i}^{\mathrm{obs}}-C_{i}^{\mathrm{AS}})$, with $\varepsilon_{0}=230~\mu$m \citep{Frayer2018}. We estimate the aperture efficiency at a wavelength of $3$~mm, using an aperture efficiency of $0.71$ at $21$~cm. The estimated aperture efficiencies are presented in Figure~\ref{fig:eta}. On average the aperture efficiency drops by $10\pm5\%$, with a maximum drop of $18\%$ (from $0.28$ to $0.23$).

During the acquisition of the point clouds used for this analysis the wind speed was always lower than $4$~m~s$^{-1}$ (Figure~\ref{fig:external}), the displacements of the feed arm relative to the primary reflector were in the range $\pm2^{\prime\prime}$, and the acceleration of the receiver cabin was $\pm2\times10^{-4}$g (Figure~\ref{fig:accqd}). These results suggest that using the LASSI under similar, or better, observing conditions it should be possible to improve the aperture efficiency of the telescope relative to not applying any corrections in the form of thermal Zernike polynomials.

We also explore how averaging the point cloud acquired when the AS is in its reference state affects the quality of the deformation maps and the measured deformations. To average the point clouds we use a weighted mean, with the rms of the residuals between the point cloud and the best fit paraboloid as weights. When using the averaged point cloud we have $\Delta C_{i}=10\pm78~\mu$m and $\sigma_{\mathrm{LASSI}}=73\pm25~\mu$m, consistent with the value obtained using individual point clouds. This suggest that stochastic effects are not limiting the accuracy of the LASSI to measure deformations.

Finally, we study how the accuracy changes with time. For this, we use a single point cloud as reference when computing deformation maps. If we use the first reference scan throughout the night, then $\sigma_{\mathrm{LASSI}}$ increases to $145\pm30~\mu$m, and the aperture efficiency at $3$~mm drops by $31\pm12\%$ on average (from $0.28$ to $0.19$). In this case we note that $\sigma_{\mathrm{LASSI}}$ increases with time, suggesting that either the surface of the primary reflector or the TLS is changing over this time. Given that the temperature changed from $1$~C to $-0.5$~C during the observations, what was changing is quite sensitive to temperature variations. If we use a point cloud acquired at the middle of the night, then $\sigma_{\mathrm{LASSI}}=82\pm19~\mu$m, consistent with the value obtained using point clouds acquired consecutively to compute the deformation maps.

\section{Discussion}

\subsection{Improvements to LASSI}

For the experiments presented in this work the orientation of the TLS was fixed, hence all the scans were obtained at the same elevation. In the future the TLS will be installed in a mount that includes a hinge, thus enabling it to keep its orientation within its operational range at any elevation. With this hinged mount it will be possible to scan the primary reflector at any elevation. In order to minimize the possible additional movement of the TLS due to this additional degree of freedom the new mount will include a brake that will be activated while the TLS is scanning.

Here we have investigated the use of differential measurements to reduce the systematics introduced by the TLS. However, it is also possible to calibrate part of the misalignment of the optics of the TLS, either in a laboratory\citep{Muralikrishnan2015}, in a dedicated range\citep{Medic2020}, or using a surface with a known shape (e.g., the primary reflector of the telescope) as calibration surface.\citep{Holst2017,Holst2019} Each of these methods has its own costs and limitations. For the LASSI we will start by investigating the use of the primary reflector as a calibration surface, since implementing this calibration method would require no additional costs, it would benefit from the high surface accuracy of the primary reflector of the GBT, and it would allow to calibrate the TLS in-situ, without the additional overheads (e.g., moving the TLS and installing it at a new location). To carry out this calibration scheme it is necessary to scan the surface of the primary reflector at a range of elevations, thus it will require the new hinged mount. One possible limitation of this approach is that so far it has not been shown if this type of calibration is repeatable in time. We will explore this approach once the hinged mount becomes available.

For the scans presented here we used a temporary mount, and temporary connections. Once the scanner is permanently connected it will be connected to a network with a higher speed, which will reduce the time required to transfer the point cloud to the analysis server. Additionally, the current smoothing is performed on a single GPU. In the future we will take advantage of the multiple GPUs available as part of the Versatile GBT Astronomical Spectrometer (VEGAS) \citep{Prestage2015} to process the point clouds using multiple GPUs in parallel. These changes will reduce the time required to process a point cloud after it has been captured.

\subsection{Future experiments}

So far we have used observations during two nights to characterize the performance of the TLS and our algorithms for measuring deformations. In the future, once the TLS is permanently installed on the GBT, it will be possible to carry out observations more frequently, as it will not be necessary to mount the TLS each time. When measuring deformations we only considered three different Zernike polynomials, independently, $Z_{4}$, $Z_{7}$ and $Z_{13}$. The first two are examples of low order Zernike polynomials with no azimuthal symmetry, while the latter is a polynomial with no azimuthal dependency. Even though these are representative cases, they are only a subset of the Zernike polynomials required to bring the surface error of the GBT's primary reflector to $230~\mu$m. In general, the corrections to the surface of the primary reflector derived from OOF holography consist of $15$ to $21$ Zernike polynomials. In the future we will conduct observations to determine the accuracy with which we can measure the missing Zernike polynomials individually and including all of them at once. Particular emphasis will be put into measuring the low order Zernike polynomials, $Z_{4}$, $Z_{5}$ and $Z_{6}$.

It is also necessary to characterize how the systematics from TLS change with time, particularly how the misalignment of its optics changes with temperature \citep{Medic2021,Janssen2021}. This will require repeated measurements of the calibration parameters for the TLS under different operating temperatures. These experiments will also help understand if the assumptions behind Eq.~\ref{eq:defmap} hold over longer time scales. Particularly, the assumption that the scanner's systematics should cancel out when taking the difference between two point clouds.

\section{Summary}\label{summary}

We have tested the potential of using a commercial off-the-shelf TLS to measure thermal deformations over the $100~$m parabolic reflector of the GBT, the Laser Antenna Surface Scanning Instrument (LASSI). Using the AS on the primary reflector of the GBT we simulated thermal deformations. The point clouds produced by the TLS were processed for deformation analysis. Since commercial TLS are not rated to measure deformations of a hundreds of micrometers, particularly because their optics are not perfectly aligned, we used the difference between point clouds to remove the systematic effects introduced by the imperfect optics of the TLS. We find that with our processing algorithms and the use of the difference between point clouds it is possible to accurately measure deformations of the primary reflector of the GBT.

The difference between the input and measured deformations is less than $\pm140~\mu$m, for deformations in the form of Zernike polynomials. These differences result in a wavefront error of $74\pm20~\mu$m, which we estimate would reduce the aperture efficiency of the GBT at $3$~mm by $<20\%$. A preliminary analysis of the influence of external factors on the accuracy of these measurements suggests that the wind speed at the GBT is the dominant factor. This effect could be explained by the motion of the feed arm relative to the primary reflector induced by the wind. During the experiments used to measure deformations the wind speeds were $<4$~m~s$^{-1}$.

We also described future improvements to the system, and future experiments necessary to understand if the LASSI is a viable method to accurately measure thermal deformations of the primary reflector of the GBT during observations at short wavelengths ($\lambda\leq1.5$~cm).

\section*{Acknowledgments}
We thank C.~Holst for sharing his expertize about terrestrial laser scanning, and the annonymous referees for their comments which improved the manuscript.
LASSI ("Enhancing GBT Metrology to support high resolution $3$~mm molecular imaging for the U.S. Community") is supported by the National Science Foundation under Award Number AST-1836009. 
Green Bank Observatory is supported by the National Science Foundation and is operated by Associated Universities, Inc.

\bibliography{lassi}

\end{document}